\documentclass[aps,prb,reprint,superscriptaddress]{revtex4-2}

\usepackage{graphicx}% Include figure files
\usepackage{dcolumn}% Align table columns on decimal point
\usepackage{bm}% bold math
\usepackage{textcomp} %Added to address error
\usepackage{amsmath}
\usepackage{xcolor}
\usepackage{xcolor}

\begin{document}

%\preprint{APS/123-QED}

\title{Integrating Bayesian Inference with Scanning Probe Experiments for Robust Identification of Surface Adsorbate Configurations}

\author{Jari J\"{a}rvi}
\email[]{jari.jarvi@aalto.fi}
\author{Benjamin Alldritt}
\author{Ond\v{r}ej Krej\v{c}\'{i}}
\affiliation{Department of Applied Physics, Aalto University, P.O. Box 11100, FI-00076 AALTO, Finland}
\author{Milica Todorovi\'{c}}
\affiliation{Department of Applied Physics, Aalto University, P.O. Box 11100, FI-00076 AALTO, Finland}
\affiliation{Department of Mechanical and Materials Engineering, University of Turku, Vesilinnantie 5, Turku, FI-20014, Finland}
\author{Peter Liljeroth}
\author{Patrick Rinke}
\affiliation{Department of Applied Physics, Aalto University, P.O. Box 11100, FI-00076 AALTO, Finland}

\begin{abstract}

Controlling the properties of organic/inorganic materials requires detailed knowledge of their molecular adsorption geometries. This is often unattainable, even with current state-of-the-art tools. Visualizing the structure of complex non-planar adsorbates with atomic force microscopy (AFM) is challenging, and identifying it computationally is intractable with conventional structure search. In this fresh approach, cross-disciplinary tools are integrated for a robust and automated identification of 3D adsorbate configurations. Bayesian optimization is employed with first-principles simulations for accurate and unbiased structure inference of multiple adsorbates. The corresponding AFM simulations then allow fingerprinting adsorbate structures that appear in AFM experimental images. In the instance of bulky (1$S$)-camphor adsorbed on the Cu(111) surface, three matching AFM image contrasts are found, which allow correlating experimental image features to distinct cases of molecular adsorption.

\end{abstract}

\maketitle

%%%%%%%%%%%%%%%%%%%%%%%%%%%%%%%%%%%%%%%%%%%%%%%%%%%%%%%%%%%%%%%%%%%%%%%%%%%%%%%%
\section{\label{sec:intro} Introduction}

The adsorption geometry of molecular adsorbates is a key parameter controlling many on-surface properties, such as diffusion and, more generally, the mechanism and yield of heterogeneous chemical reactions \cite{Rotter2016,Brandt2009}. In the field of heterogeneous catalysis, powerful electron-microscopy-based techniques are now capable of resolving the structure of the catalyst surface on the atomic scale \cite{Gao2017_review,Boyes2020_review}. However, these methods still cannot determine the adsorption configuration of the reactant on the active site. In general, visualizing non-planar adsorption structures on the single-molecule level remains a challenging task.

The current state-of-the-art in visualizing nanostructures with atomic resolution is scanning probe microscopy. Atomic resolution can be achieved with non-contact atomic force microscopy (AFM) with functionalized carbon monoxide (CO) tips \cite{Bartels1997,Gross1110}. CO-AFM excels in structure analysis of planar organic molecules in real space, facilitating the direct identification of molecular structures \cite{Gross1110} and conformations \cite{PhysRevLett.108.086101}.  Existing work has primarily focused on geometrically flat (planar) species. Only a few 3-dimensional (i.e. non-planar) molecules with limited conformations have been investigated \cite{PhysRevLett.108.086101, Albrecht2013, Albrecht2015, Albrecht2016, Jarvis2015, Gross_sci_C60_2012, Kawai2017, Schulz2017, Kawai2018, Pawlak2019, Moreno2015}, as the interpretation of different 3D adsorbate conformations remains a considerable challenge.

First principles calculations, e.g. density-functional theory (DFT) \cite{Hohenberg1964, Kohn1965}, are a powerful tool for simulating and identifying adsorption structures. DFT provides an accurate quantum mechanical description of important adsorbate-surface interactions, but exhaustive structure search is needed to determine all the different adsorbate structures. AFM images of 2-dimensional adsorbates can inform the structure search about the molecular registry and orientation at the substrate \cite{Todorovic2018, Schuler2015, Kawai2018}. In contrast, images of complex non-planar molecules are often not conclusive enough, and estimating the structures using chemical intuition is difficult. Here, we propose a computationally efficient method to determine the structure of 3-dimensional organic adsorbates using Bayesian inference with chemical building blocks and AFM simulations.

\begin{figure}
\includegraphics{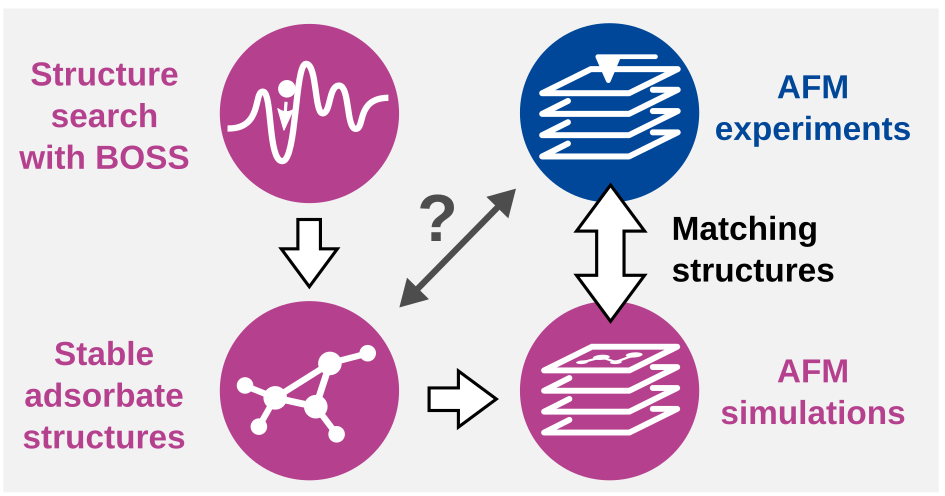}
\caption{\label{fig:intro}\textbf{Concepts and workflow of the proposed methodology.} Identifying the structure of 3D adsorbates is often difficult from AFM experimental images (black arrow). In our combined approach, we first perform global structure search with the Bayesian Optimization Structure Search (BOSS) method and density-functional theory (DFT) to identify the stable model structures. We then simulate atomic force microscopy (AFM) images for the identified structures. We analyze the features in the simulated images and compare them to the corresponding features in experimental AFM images to detect matching configurations (white arrows).}
\end{figure}

\begin{figure*}
\includegraphics{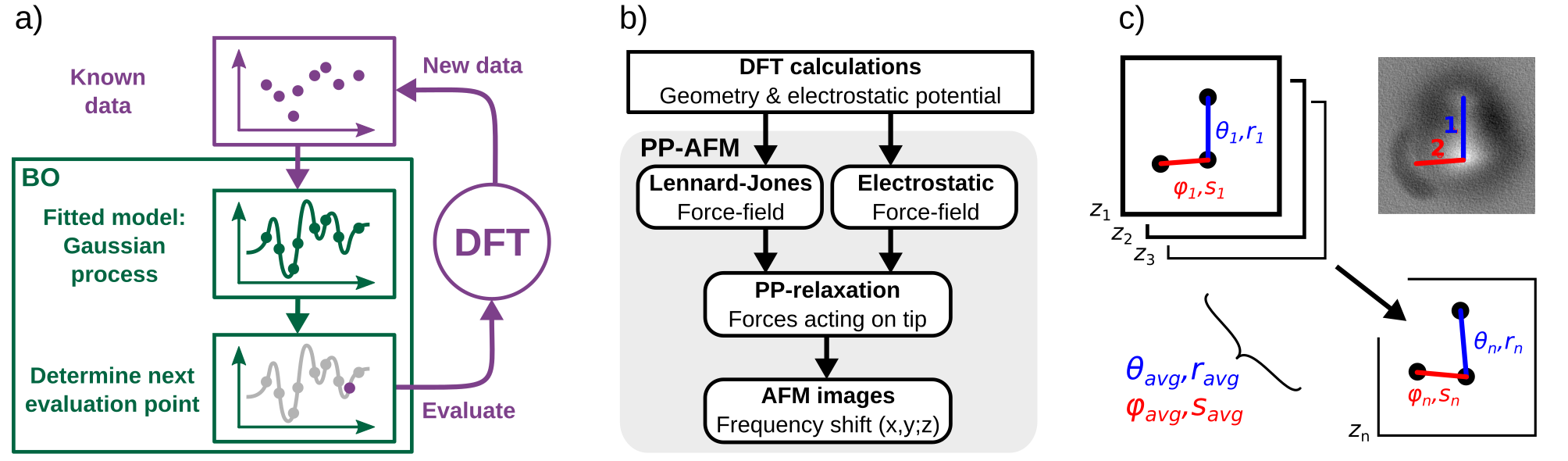}
\caption{\label{fig:methods}\textbf{Methodology for our integrated approach.} a) Basic principle of the BOSS method, in which Bayesian optimization (BO) is applied iteratively with DFT to build a surrogate model of the AES. In BO, the known data is first fitted in a Gaussian process, after which the next evaluation point is determined using an acquisition function. The new point is evaluated with DFT and the process is repeated with the new data included. b) Workflow of the Probe Particle (PP)-AFM simulation method. The geometry and electrostatic potential of the structure from DFT are used to compute molecular mechanic force-fields. The PP, which mimics the flexible tip-apex, relaxes in this force-field. The final force acting on the last metallic (fixed) atom of the tip is used to calculate the frequency shift $\Delta f$. c) Experiment-simulation image matching, in which AFM images are analyzed via orientations ($\theta, \varphi$) and lengths ($r, s$) of the observed  bright elongated features (BEFs) 1 and 2. The analysis is performed on a stack of $n$ images, obtained at different heights of the CO tip. The orientation and length of each BEF is calculated as an average of the measured values in the image stack.}
\end{figure*}

Stable adsorbate structures can be objectively identified as the local minima of the adsorption energy surface (AES). Thorough sampling of high-dimensional AESs with conventional methods \cite{Goedecker2004, Li2012} requires excessively many energy calculations, constraining us to fast force field methods which do not have the required accuracy to describe molecular adsorption. To overcome these limitations, novel Bayesian inference methods have recently been employed \cite{Packwood2017, Carr2016}. Gaussian process regression \cite{Rasmussen2006} is a particularly promising technique capable of constructing a surrogate model of the AES with a modest number of energy points. When combined with active learning in Bayesian optimization (BO) \cite{Shahriari2016}, it can be used to accelerate the construction of the AES model via strategic sampling. The complete AES then allows us to identify all the stable structures and estimate their mobility via the associated energy barriers. In this study, we rely on the recently developed Bayesian Optimization Structure Search (BOSS) method \cite{Todorovic2019,Egger2020,Fang2021,BOSSwebsite} to model the surrogate AES.

Our objective here is to construct and test new methodology for automated and robust search of adsorption geometries for bulky 3D molecules. We integrate tools from different research fields to identify adsorbate structures without any requirement of previous knowledge about the studied material. Our work flow features i) global structure search with BOSS and DFT, ii) AFM image simulation with the Probe Particle (PP)-AFM model \cite{Hapala_PRB_2014,Hapala_PRL_2014,PPwebsite}, and iii) AFM experiments (\textbf{Figure~\ref{fig:intro}}). In previous research \cite{Todorovic2019}, BOSS was applied to identify the preferred adsorption of a C$_{60}$ molecule on a TiO$_2$ anatase surface, but the comparison of global minimum models to experimental AFM images was inconclusive. Here, we extend this methodology and extract all the stable structures (local minima), which we compare to multiple different experimental configurations for a more robust test of the methodology. With this method, several experimental structures could be identified solely based on a single model of the AES. We demonstrate the success and efficiency of this approach by identifying the stable adsorbate structures of (1$S$)-camphor (C$_{10}$H$_{16}$O) on the Cu(111) surface.

Previous AFM experiments \cite{Alldritt2020} have shown that (1$S$)-camphor (a typical bulky molecule) adsorbs to Cu(111) in  different stable configurations. The adsorption structures, in particular the orientations of the molecule, were difficult to interpret from the AFM images. We use BOSS (\textbf{Figure~\ref{fig:methods}}a) to identify all the stable molecular adsorbate structures and their energy barriers. We select the most promising structures and generate simulated AFM images (Figure~\ref{fig:methods}b) for them. By correlating the features in experimental and simulated AFM images (Figure~\ref{fig:methods}c), we detect matches to identify several adsorbate structures observed in experiments.

%%%%%%%%%%%%%%%%%%%%%%%%%%%%%%%%%%%%%%%%%%%%%%%%%%%%%%%%%%%%%%%%%%%%%%%%%%%%%%%%
\section{\label{sec:res} Results}

\subsection{\label{sec:res_exp_afm} Experimental AFM images}

\begin{figure*}
\includegraphics{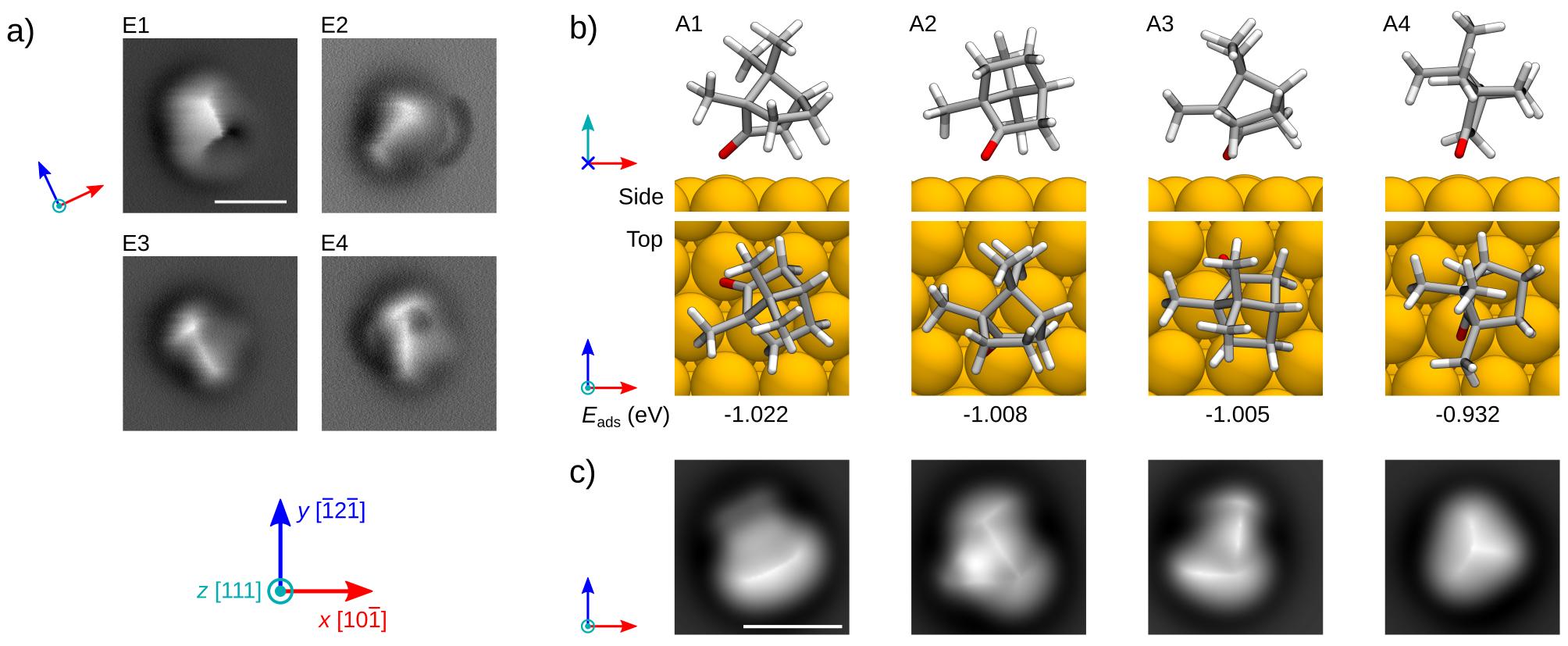} 
\caption{\label{fig:results}\textbf{Summary of key results for stable adsorbate structures of (1$S$)-camphor on Cu(111).} a) Constant-height AFM images, showing 4 different adsorbate structures. b) Stable model structures A1--A4 and their adsorption energies ($E_\text{ads}$), predicted by BOSS and relaxed with DFT. The top views of the structures are showing an area of 6.85$\times$6.85~\AA{}. c) Simulated AFM images of the model structures. Shown is a single image from the image stack, taken at height 5.6~\AA{} above the highest atom of the calculated structure. Coordinate axes indicate the Cu(111) lattice orientation. Scale bars in a) and c) are 5~\AA.}
\end{figure*}

Experiments were performed in ultra-high vacuum (UHV) on a clean Cu(111) surface, which corresponds well to a defect-free computational model. (1$S$)-camphor was deposited onto a Cu(111) surface held at $T = 20$ K. The low temperature reduced the mobility of the molecules on the substrate,  but did not prevent them from sampling various conformations during  adsorption. After deposition, we imaged a random selection of deposited molecules that were away from step edges and tip preparation areas. We collected 14 images of adsorbed (1$S$)-camphor molecules by CO-AFM (\textbf{Figure S1} in the Supporting Information (SI)) and found the molecule adsorbed in multiple configurations.

Approximately half of the imaged structures featured a mobile adsorbate, in which the orientation of (1$S$)-camphor changed during the measurement. In this study, we exclude these mobile adsorbates, as well as 5 static structures where AFM amplitude instabilities occurred at close tip-sample distances (details provided in the SI). We focus on 4 static structures depicted in images E1--E4 (\textbf{Figure~\ref{fig:results}}a and \textbf{Figures S2--S5}), which are the most strongly adsorbed and the least mobile on this substrate. To characterise each one, we collected detailed image stacks at different tip-surface heights (8 to 11 images per structure).

CO-AFM images of camphor contain bulky oval shapes, with several linear bright features in the centre. Such indistinct image types are typical in AFM imaging of 3D objects \cite{Kawai2018} and, unlike AFM images of 2D molecules, do not lend themselves to easy interpretation. To facilitate structure identification, we analyzed the distinctive features that appear in the images of each structure, taking into account entire stacks of experimental images collected. The orientation of (1$S$)-camphor is evaluated with respect to the nearest crystallographic axis of the Cu(111) surface lattice in the clockwise direction, in the range $[0, 60]^\circ$. For the angle determination, we note that the Cu lattice is rotated by $25 \pm 1^\circ$ in the counter-clockwise direction compared to the lattice in the computational model. The lattice orientation was confirmed with two separate measurements --- with a clean Cu surface and with an adsorbed (1$S$)-camphor molecule on the surface.

\subsection{\label{sec:res_boss} Identifying stable adsorbate configurations}

\begin{table*}
\caption{\label{tab:res} \textbf{Comparison of model structures and AFM images with BEFs.} Adsorption energy ($E_\text{ads}$) of (1$S$)-camphor on Cu(111), energy barriers of molecular rotation ($E_\text{R}$) and diffusion($E_\text{D}$), distance ($d_\text{DFT}$) and orientation ($\theta_\text{DFT}$) between the two topmost atoms of (1$S$)-camphor on Cu(111) in the model structures predicted by BOSS. Average length ($\overline{d}$) and orientation ($\overline{\theta}$) of the main BEFs in the stack of simulated (sim) and experimental (exp) AFM images. The average lengths and orientations of all BEFs and their standard deviations are provided in the SI.}
\begin{ruledtabular}
\begin{tabular}{cccccc@{\hskip 3em}cc@{\hskip 3em}ccc}
& \multicolumn{5}{c}{BOSS/DFT}
& \multicolumn{2}{l}{simulated AFM}
&& \multicolumn{2}{l}{experimental AFM} \\
& $E_\text{ads}$ [eV] & $E_\text{R}$ [eV] & $E_\text{D}$ [eV] & $d_\text{DFT}$ [\AA] & $\theta_\text{DFT}$ [$^\circ$]  & $\overline{d}_\text{sim}$ [\AA] & $\overline{\theta}_\text{sim}$ [$^\circ$] && $\overline{d}_\text{exp}$ [\AA] & $\overline{\theta}_\text{exp}$ [$^\circ$] \\
\hline
A1 & -1.022 & 0.232 & 0.045 & 1.77 & 25.2 & 2.76 & 27.9 & E1 & 2.74 & 7.8 \\
A2 & -1.008 & 0.216 & 0.034 & 2.23 & 3.9 & 3.82 & 3.0 & E2 & 3.55 & 27.4 \\
A3 & -1.005 & 0.183 & 0.008 & 2.54 & 24.3 & 4.04 & 29.1 & E3 & 3.77 & 38.1 \\
A4 & -0.932 & 0.278 & 0.027 & 1.77 & 19.6 & 3.16 & 23.7 & E4 & 4.00 & 13.8 \\
\end{tabular}
\end{ruledtabular}
\end{table*}

In a preparatory study \cite{Jarvi2020a}, we applied BOSS with DFT to identify the stable adsorbate structures of (1$S$)-camphor on the Cu(111) surface. We identified 8 stable structures with varying molecular orientations, adsorption sites and energy barriers. Based on their adsorption properties, the structures were classified into two categories, A and B (Ox and Hy in previous study \cite{Jarvi2020a}). Class A structures, in which (1$S$)-camphor chemisorbs to Cu(111) via oxygen (O), are the most stable and have the highest energy barriers of molecular rotation and diffusion. In class B, (1$S$)-camphor physisorbs to Cu(111) via hydrocarbon interactions.

Here, we analyse the observed configurations and select the model structures that most closely correspond to the molecules that were exhaustively characterised in experiments. During AFM imaging, the adsorbates that underwent rotations or translations in response to tip approach were excluded from further considerations. We therefore disregard class B model structures and A5, which have very low barriers to rotation and diffusion. Structures A1--A4 (Figure~\ref{fig:results}b and \textbf{Table~\ref{tab:res}}) are the least mobile and most strongly adsorbed, and they make the best candidates for the static adsorbates in experiments.

To ensure that the level of \emph{ab initio} theory employed does not affect our conclusions, we verified the accuracy of structures A1--A4 by: i) representing vdW interactions with many-body dispersion \cite{Tkatchenko2012} instead of the semi-empirical Tkatchenko-Scheffler (TS) method, and ii) applying the HSE hybrid exchange-correlation functional \cite{Heyd2003} instead of semi-local PBE \cite{Perdew1996}. We observed negligible changes in the adsorption geometries (\textbf{Tables S3 and S4} in the SI), confirming that PBE with TS dispersion is sufficiently accurate to describe these materials. Next, we employ these structures to generate simulated AFM images, which represent the most stable adsorbates of (1$S$)-camphor on Cu(111).

\subsection{\label{sec:res_sim_afm} Simulating AFM images}

With the identified stable adsorbates, we produced simulated AFM images (Figure~\ref{fig:results}c) for direct comparison of the structures with AFM experiments. We simulated CO-AFM with the PP-AFM method (Figure~\ref{fig:methods}b) \cite{Hapala_PRB_2014,Hapala_PRL_2014,PPwebsite} using different heights of the CO tip from the surface. Our simulations of CO-AFM images for structures A1--A4 are in the height range [5.3, 6.5]~\AA, measured as the distance between the CO tip and the highest atom of (1$S$)-camphor. The images were produced at height steps of 0.1~\AA, which provides a discernible difference between images at each step.

For each structure, we obtained a stack of 11 images, from which we then extract distinct features at different heights (Figure~\ref{fig:results}c and \textbf{Figures S6--S9} in the SI). These features start to appear as bright spots above the highest protruding atoms of the simulated structure in the top-most images. As the scan is moving lower -- towards the molecules -- they begin to elongate. These bright elongated features (BEFs), as we call them, emerge over and between the bright spots of the top-most atoms \cite{Hapala_PRB_2014}. Their appearance and number can significantly change with the scan height. However, most of them remain recognizable to the lower limit of our scans. We choose the most pronounced and well defined BEFs as the key fingerprint features for matching simulated and experimental AFM images, as discussed below in Section~\ref{sec:matching}.

\subsection{\label{sec:matching} Detecting matching structures}

\begin{figure*}
\includegraphics{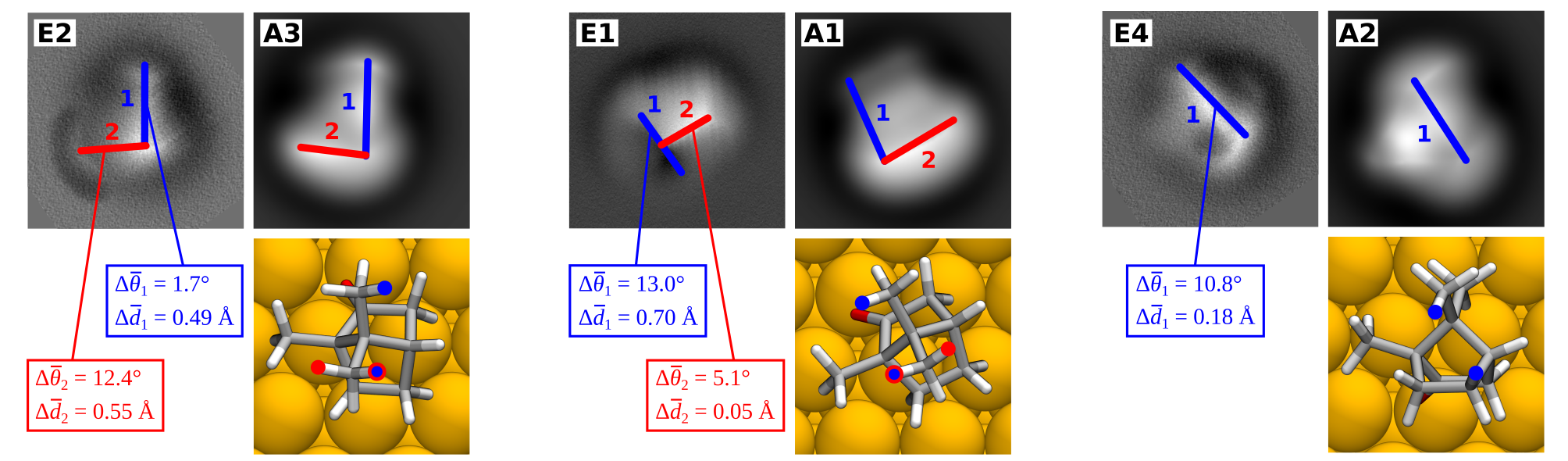}
\caption{\label{fig:matches}\textbf{Detected matches between experimental and simulated structures.} Matches between the experimental (E) and simulated (A) structures E2-A3, E1-A1, and E4-A2 are compared via the orientations and lengths of the identified BEFs 1 (blue) and 2 (red) in the AFM images. The top view of each simulated structure shows the topmost atoms (blue and red), which are the origin of the BEFs. Matching accuracy is evaluated via the difference in the average orientations ($\Delta \overline{\theta}$) and lengths ($\Delta \overline{d}$) of the observed BEFs between the experimental and simulated images.}
\end{figure*}

We combine our analysis of the observed BEFs in the simulated and experimental AFM images and detect matches to identify the structures observed in AFM experiments. With each structure, we identify the orientations and lengths of the BEFs in each image in the stack (Figure~\ref{fig:methods}c). This is done by finding the local maxima in the BEFs and connecting them with a straight line. Comparing the length and orientation of BEFs in CO-AFM is standard for imagining planar molecules. There the BEFs look like bright lines and are often called the apparent bonds, e.g. in \cite{Gross_sci_C60_2012,Hapala_NC_2016}. The 3D nature of the (1$S$)-camphor molecule precludes a similar matching strategy, because the scan height cannot easily be inferred from the image contrast. Instead a new matching strategy is required. Due to their variation in the orientations and lengths at different heights, we compare the BEFs via their average orientation and length over the stack of images. This way we ensure that small deviations of the scan heights are not affecting the overall results. The orientations are measured as the angle from the nearest crystallographic axis of the Cu(111) surface in the clockwise direction, in the range [0, 60]$^\circ$. We measure the lengths as the distance between the BEF maxima. Images for all heights and the average lengths and orientations of all analyzed features and their standard deviations are provided in the SI.

In the simulated AFM images, the lengths of the main BEFs vary from 2.76 to 4.04~\AA~(Table~\ref{tab:res}). Their orientations show two distinct groups, in which structure A2 is nearly parallel (3.0$^\circ$) to the crystallographic axis of the Cu(111) surface, and the other structures are near the middle region between the axes (on average 26.9$^\circ$). In the experimental images, the main BEF of structure E1 has noticeably shorter length (2.74~\AA) than the other structures (on average 3.8~\AA). Their orientations vary from 7.8 to 38.1$^\circ$, with no distinct grouping in preferred directions.

The standard deviations of molecular orientations in the stacks of images are as large as $5^\circ$ \textbf{(Tables~SI and SII)}. The length of the BEFs in most cases increases by 0.1~\AA~for every~\AA~that the tip is approaching the molecule.  Consequently, the standard deviations of the BEF lengths are large, up to 0.8 and 0.5~\AA~in the simulated and experimental images, respectively. 

We match the structures using two identified BEFs in structures A1, A3, E1, and E2, and one BEF in structures A2 and E4. In this analysis, we detect 3 possible matches between structures E2-A3, E1-A1, and E4-A2 (\textbf{Figure~\ref{fig:matches}}). In E2 and A3, the orientations of the main and secondary BEFs agree within 1.7 and 12.4$^\circ$, respectively. The corresponding lengths agree within 0.5 and 0.6~\AA. In E1 and A1, the orientations of the BEFs agree within 13.0 and 5.1$^\circ$, and the lengths within 0.7 and 0.1~\AA, respectively. In E4 and A2, we compare a single BEF, in which the orientation agrees within 10.8$^\circ$ and the length within 0.2~\AA.

We also analyze how the BEFs in the simulated AFM images correspond to the atomistic model structures. For this, we measure the distance and orientation between the two topmost atoms of (1$S$)-camphor in the model structures A1--A4. Here, we observe two distinct groups of distances and orientations (Table~\ref{tab:res}). In structures A1 and A4, the distance of 1.77~\AA~corresponds to the separation of H atoms in the same methyl group.  Distances in structures A2 and A3 are considerably longer (2.23 and 2.54~\AA, respectively) and originate from H atoms in different groups. The orientations exhibit two preferred directions: nearly parallel to the crystallographic axis (structure A2, $3.9^\circ$), and in the middle region between neighboring axes (structures A1, A3, and A4, on average $23.0^\circ$). The distance between the top atoms in the model structures is 1.4~\AA~shorter, on average, than the corresponding BEF length in simulated AFM images. The average difference in the orientation between the model structures and simulated images is $3.1^\circ$, without a clear trend in either rotation direction.

%%%%%%%%%%%%%%%%%%%%%%%%%%%%%%%%%%%%%%%%%%%%%%%%%%%%%%%%%%%%%%%%%%%%%%%%%%%%%%%%

\section{\label{sec:disc} Discussion}

The stable adsorbate structures, which we identified with BOSS and DFT, show that (1$S$)-camphor can adsorb to Cu(111) in multiple stable configurations with varying molecular orientations and adsorption sites. This explains the different types of adsorbates observed in AFM experiments. To interpret the experiments, we generated simulated AFM images of the most stable and least mobile model structures for a direct comparison with experiments using the BEFs. This workflow can be generally used for comparison of adsorbed 3D molecules in  CO-AFM, with the possibility to quantify the quality of the match. 

In this comparison, we observed very similar features between experimental and simulated AFM images. The primary criterion for a good match is the agreement of the BEFs orientation angles, while the BEF length comparison is a secondary consideration. Feature orientations with respect to the substrate have lower error bars than feature lengths and are a more reliable indicator of underlying structures. We immediately detected 3 good matches between structures E2-A3, E1-A1, and E4-A2, in which the orientations and lengths of the observed BEFs are in good agreement. In these matches, we also took into account the deviation of the BEFs in each image stack.

The best match is between structures E2 and A3, in which the two analyzed BEFs agree closely between simulations and experiments. Similarly good agreement of two BEFs was found between structures E1 and A1. Here, however, the BEFs emerge in different order as the CO tip is approaching the molecule. This can be explained by a minor tilt in the orientation of the molecule, which can be induced by its interaction with the tip, as discussed previously \cite{Alldritt2020}. In the third match, between E4 and A2, the length of the single analyzed BEF agrees closely and the orientation is only slightly outside the specified error threshold (within $11^\circ$).

The adsorption of (1$S$)-camphor on Cu(111) has been previously studied by Alldritt~et~al. \cite{Alldritt2020} using an artificial neural network (ANN) with image descriptors for automated structure discovery. By comparing the image descriptors for systematically rotated isolated molecules against those of experimental images, they identified the most likely molecular orientation. The ANN predicted that, based on the AFM images, (1$S$)-camphor binds to the Cu surface via hydrocarbon interactions. We found this physisorbed structure to be a local minimum, with chemisorption via the O-Cu bond to be energetically more favorable \cite{Jarvi2020a}. The contrasting results can be explained by a fundamental difference in the two machine learning approaches. ANN-led structure discovery aims to extract most likely 3D molecular structures behind AFM images, free of surface considerations or energetics. In contrast, our BO-led structure search strives to learn the molecule-surface interaction and find all minima, which are then compared to experiment.  Alldritt~et~al. \cite{Alldritt2020} did consider surface effects in the supplementary material by optimising 500 randomly chosen molecular orientations on the surface. This produced several local minima which were different from the ANN predictions but in agreement with the stable structures in this work. 

Independent structure identification is important because the interpretation of experimental AFM images of bulky 3D structures is complicated. In contrast to the 2D adsorbate case, 3D adsorbate AFM images contain BEFs whose measured orientations and lengths exhibit much larger differences between theory and experiment, and thus require a thorough statistical analysis. We also note that the trend of increasing BEF length with tip approach is opposite to what was reported for 2D molecules in full monolayer \cite{Hapala_NC_2016} or borders of polygons of C$_{60}$ molecules \cite{Gross_sci_C60_2012}. 

To further clarify the experimental AFM images, we have taken great care with the AFM simulation approach. With 3D molecules the overall match and especially the visual comparison between simulated and experimental images is considerably more intricate than with  planar molecules or 2D materials. AFM is extremely sensitive to the z-coordinate of the atom position and even very minor changes will affect the image contrast in a noticeable way. Even for planar molecules it was shown that the apparent length of the BEFs in simulated AFM differs from the experimental measurements and there is also scatter in the experimental values for planar molecules~\cite{Gross_sci_C60_2012,Hapala_PRB_2014,Hapala_NC_2016}. 

The AFM simulation model can be made more sophisticated by adjusting the electrostatic potential and the Pauli repulsion ~\cite{Ellner_NL_2016,Ellner_ACSNano_2019}, but we have found the PP-AFM model to be adequate for this study as the differences between simulated and experimental images are likely to arise from minute differences in the molecular geometries. Structural relaxations of the adsorbate at the very small tip-molecule distances can also affect the image contrast. In the future, the quality of structure matching could be further enhanced by implementing sample response to the presence of tip in the PP-AFM model. Nonetheless, the energetic stability of the structures identified here, as well as their high rotational and translation barriers, strengthen the proposed matches between experiments and simulations.

\section{\label{sec:conc} Conclusion}

In conclusion, we have proposed a new approach to investigate the structure of complex 3D adsorbates. We have integrated a set of tools from different fields, using Bayesian inference enhanced structure search, AFM simulations with the PP-AFM model, and CO-AFM experiments. With BOSS, we constructed a surrogate model of the complete AES to extract the stable model structures and their energy barriers of molecular mobility. This allowed us to infer different adsorbate types independently of AFM images, and free of chemical intuition. PP-AFM simulated images then facilitated a direct comparison of the model structures with CO-AFM experiments. The combination of findings derived from different sources is key to robust identification of distinct adsorbate geometries in experimental images. In the case of (1$S$)-camphor on the Cu(111) surface, we identified three different adsorbate geometries in the otherwise incomprehensible features of AFM experimental images. This Bayesian-based general approach can be applied to other adsorption structure search problems and combined with other experimental techniques. Uncovering the complete adsorption geometry of 3D adsorbates at the single molecule level is the key towards a detailed control of surface structure and properties and to the understanding of reaction products, intermediates and pathways of on-surface chemical reactions.

%%%%%%%%%%%%%%%%%%%%%%%%%%%%%%%%%%%%%%%%%%%%%%%%%%%%%%%%%%%%%%%%%%%%%%%%%%%%%%%%
\section{\label{sec:meth} Methods} 

\subsection{Experimental AFM} A polished Cu(111) single-crystal (Mateck/Germany) was prepared by repeated Ne+ sputtering (0.75 keV, 15 mA, 20 min) and annealing (850-900 K, 5 min) cycles. Sample temperatures during annealing were measured with a pyrometer (SensorTherm Metis MI16). Following the cleaning process, the Cu(111) surface was verified by scanning tunneling microscopy (STM), investigating impurity concentration and terrace size.

A high-purity gas line with leak valve was prepared for deposition of the (1$S$)-camphor (Sigma-Aldrich, purity $>$ 98.5$\%$) molecules directly into the STM chamber. The gas line was baked at 400 K for 24 hours. The (1$S$)-camphor molecules were placed in the gas line, pumped, and briefly heated to ca. 370 K before returning to room temperature. (1$S$)-camphor was introduced into the STM via the leak valve and deposited onto the Cu(111) surface held at $T=20$ K. CO gas was deposited via the same gas line onto the Cu(111) surface held at $T=20$ K. 

The STM and CO-AFM images were taken with a Createc LT-STM/AFM with a commercial qPlus sensor with a Pt/Ir tip, operating at approximately $T = 5$\ K in UHV at a pressure of $1\times10^{-10}$ mbar. The quartz cantilever (qPlus sensor) had a resonance frequency of $f_{0}=29939$ Hz, a quality factor $Q=101099$, and was operating with an oscillation amplitude $A=50$ pm. Tip conditioning was performed by repeatedly bringing the tip into contact with the copper surface and applying bias pulses until the necessary STM resolution was achieved. The tip apex was functionalized with a CO molecule before AFM measurements. The STM images were recorded in constant-current mode, while the AFM operated in constant-height mode. Raw data was used as input for the image analysis. In order to minimize experimental artefacts that would cause problems with interpretation, we have implemented the following measures: Checking the background $\Delta f$ before CO pickup (smaller value indicates sharper overall tip); scanning another CO to ensure the symmetry of the CO tip after tip passivization and prior to further AFM imaging; and confirming that the excitation (dissipation) signal remains flat/featureless during the AFM measurements.

\subsection{Bayesian Optimization Structure Search}

Global phase space exploration for molecular adsorption of (1$S$)-camphor on Cu(111) was carried out in 6D with BOSS \cite{Jarvi2020a}. We defined the AES in the search space of molecular position and orientation using 3 translational and 3 rotational degrees of freedom. 609 DFT calculations were sufficient to construct the model with applied symmetries in the orthogonal unit cell. We identified the stable structures in the AES minima and verified them with full relaxation in DFT (i.e.  unrestricted motion of all atoms). The energy barriers of molecular diffusion ($E_\text{D}$) were evaluated from the AES model, and the rotational barriers ($E_\text{R}$) were predicted with BOSS by rotating the molecule in the relaxed structures.

\subsection{Simulated AFM} The PP-AFM simulations were based on DFT calculated geometries and electrostatic potentials \cite{Hapala_PRL_2014}. For the mechanical part of PP-AFM we employed the OPLS force-field \cite{Jorgensen1988} for the Lennard-Jones interactions and a PP lateral stiffness of 0.24~N m$^{-1}$~\cite{Hapala_NC_2016}. The PP was set 3~\AA~below the last metallic atom of the tip~\cite{Ellner_NL_2016}. The electrostatic nature of the CO-tip was represented by a negative quadrupole moment on the PP with a moment of -0.025~e$\times$\AA$^2$~\cite{delaTorre_PRL_2017}. We used a peak-to-peak amplitude of 1.0~\AA~for the conversion of forces to frequency shifts $\Delta f$.  All heights refer to the center of the tip oscillations. 

We simulated a 3D stack of AFM images for heights in between 6.5 to 5.3~\AA~above the highest atom of the structures. The height is given with respect to the last metal atom of the tip. The height step between each image was 0.1~\AA. We also studied simulated images much closer than in the case of 2D molecules. This is due to hydrogen atoms that are responsible for the contrast in the AFM images. Hydrogen atoms evince Pauli repulsion much closer to the nucleus than carbons, which are important for the contrast of 2D molecules.

For each structure, we analyzed 11 images, in which we measured orientations ($\theta_\text{sim}$) and lengths ($d_\text{sim}$) of the most pronounced lines.  The Gwyddion program~\cite{Necas2012,GwyddionWebsite} was used to find local maxima for the measurements. The lines of the measurements are marked in Figures~S6--S9. We also performed statistical analysis and linear fitting on the measured data. The results are presented in Table~S2. The results for the most prominent lines (marked as 1 in Figures~S6--S9 and Table~S2) are shown in Table~\ref{tab:res}.

\subsection{Image analysis} We conducted an extensive image analysis to capture any statistical variation in the features of the experimental and simulated AFM images. Orientations and lengths of the prominent bright lines in experimental and computational datasets were determined by peak-to-peak analysis based on local maxima and minima in the AFM images. All orientation and length measurements were performed in Gwyddion \cite{Necas2012,GwyddionWebsite}. The results for the prominent features are available in Tables~S1 and S2.

\section*{Data availability}
The datasets generated during and/or analyzed during the current study are available from the corresponding author on reasonable request.

\section*{Code availability}
The software used in this study is freely available, the BOSS code in~\cite{BOSSwebsite}, the Probe Particle Model in~\cite{PPwebsite} and the Gwyddion in~\cite{GwyddionWebsite}.

\section*{Author Contributions}
J.~J. identified the stable model structures. O.~K. produced simulated AFM images of the model structures and B.~A. imaged the experimental structures with AFM. O.~K. and B.~A. carried out the image analysis. M.~T., P.~L., and P.~R. conceived the study and advised the work. All authors assisted in manuscript preparation.

\section*{Materials \& Correspondence}
Correspondence and requests for materials should be addressed to J.~J.

\section*{Supporting Information} %Please delete the Supporting Information statement if it is not applicable. Please supply Supporting Information in another file. Supporting information should not be provided in .tex format
Supporting Information is available from the author.

% Acknowledgements
\section*{Acknowledgements}
We would like to thank A.~S.~Foster, F.~Urtev, and N.~Oinonen for fruitful discussion.
O.~K. would like to thank P.~Hapala for help with the AFM simulations.

O.~K. has been supported by the European Union’s Horizon 2020 research and innovation programme under the Marie Skłodowska-Curie grant agreement No 845060. M.~T. and P.~R. have received funding from the Academy of Finland via the Artificial Intelligence for Microscopic Structure Search (AIMSS) project No.~13316601 and the Flagship programme: Finnish Center for Artificial Intelligence FCAI. P.~L. has received funding from the Academy of Finland via projects No.~311012 and 314882. J.~J. has been funded by the Emil Aaltonen Foundation.

We gratefully acknowledge CSC -- IT Center for Science, Finland, and the Aalto Science-IT project for generous computational resources.

\section*{Conflict of Interest}
The authors declare no conflict of interest.

\bibliography{references}

\end{document}